\documentclass[reprint,superscriptaddress,amsmath,amssymb,aps,prl]{revtex4-2}

\usepackage{graphicx}%
\usepackage{dcolumn}%
\usepackage{multirow}
\usepackage{amssymb}
\usepackage{comment}
\usepackage{MnSymbol,wasysym}
 \usepackage{setspace}

\usepackage[dvipsnames,table,xcdraw]{xcolor}

\usepackage{hyperref}
\hypersetup{
    colorlinks=true,
    linkcolor=Blue,
    filecolor=Blue,      
    urlcolor=Blue,
    citecolor=blue,
}

\definecolor{Gray}{gray}{0.85}
\newcolumntype{g}{>{\columncolor{Gray}}r}
\newcolumntype{w}{>{\columncolor{white}}r}

\renewcommand\vec[1]{\ensuremath\boldsymbol{#1}} 
\newcommand{\D}{\mathrm{d}} 

\begin{document}

\title{Superconductivity in three-dimensional interacting doped topological insulators}

\author{Andr\'{a}s L. Szab\'{o}}
\affiliation{Institute for Theoretical Physics, ETH Zurich, 8093 Zurich, Switzerland}

\author{Bitan Roy}
\affiliation{Department of Physics, Lehigh University, Bethlehem, Pennsylvania, 18015, USA}

\date{\today}

\begin{abstract}
Three-dimensional doped Dirac insulators foster simply connected (in both topological and trivial regimes) and annular (deep inside the topological regime) Fermi surfaces (FSs) in the normal state, and allow on-site repulsions among fermions with opposite spin ($U_1$) and parity ($U_2$) eigenvalues. From an unbiased leading-order (one-loop) renormalization group analysis, controlled by a suitable $\epsilon$ expansion, we show that this system develops a strong propensity toward the nucleation of scalar $s$-wave and odd-parity pseudoscalar $p$-wave pairings, favored by repulsive $U_1$ and $U_2$ interactions, respectively, irrespective of the underlying FS topology. Our results can be pertinent for the observed superconductivity in various doped narrow gap semiconductors, and the theoretical foundation can readily be applied to investigate similar phenomenon in various doped topological materials.  
\end{abstract}

\maketitle

\emph{Introduction}.~From the baryonic universe to topological quantum crystals, Dirac theory offers a unified description. As such electrical and thermal topological insulators (TIs) belonging to any Altland-Zirnbauer symmetry class in any dimension ($d$) can be modeled by a suitable massive Dirac Hamiltonian~\cite{TITh0, TITh1, TITh2, TITh3, TITh4, TITh5, TITh6, TITh7, TITh8, TITh9, TITh10, TITh11, TITh12, TITh13, TITh14}. The ${\bf k}\cdot {\bf p}$ Hamiltonian for their candidate materials also assumes a similar form in the low energy and long-wavelength limit~\cite{TITh15}. Fascinatingly, three-dimensional (3D) doped or intercalated TIs, (Cu/Nd/Sr)$_x$Bi$_2$Se$_3$ and Sn$_{1-x}$In$_x$Te, with a Fermi surface (FS), become a fully gapped superconductor at the low temperatures~\cite{TSC:Exp1, TSC:Exp2, TSC:Exp3, TSC:Exp4, TSC:Exp5, TSC:Exp6, TSC:Exp7, TSC:Exp8, TSC:Exp9, TSC:Exp10}. The paired state is possibly topological ($p$-wave) in nature~\cite{SCTh1, SCTh2, SCTh3, SCTh4, SCTh5}, showing a zero bias conductance peak (ZBCP), stemming from surface Majorana fermions. It is also conceivable to induce a transition between topological $p$-wave and trivial $s$-wave (devoid of ZBCP) pairings in Sn$_{1-x}$In$_x$Te. Despite such exciting experimental findings in the last decade, the phase diagram of interacting doped TIs, fostering superconducting ground states still remains hazy.

\begin{figure*}
    \centering
    \includegraphics[width=1.00\linewidth]{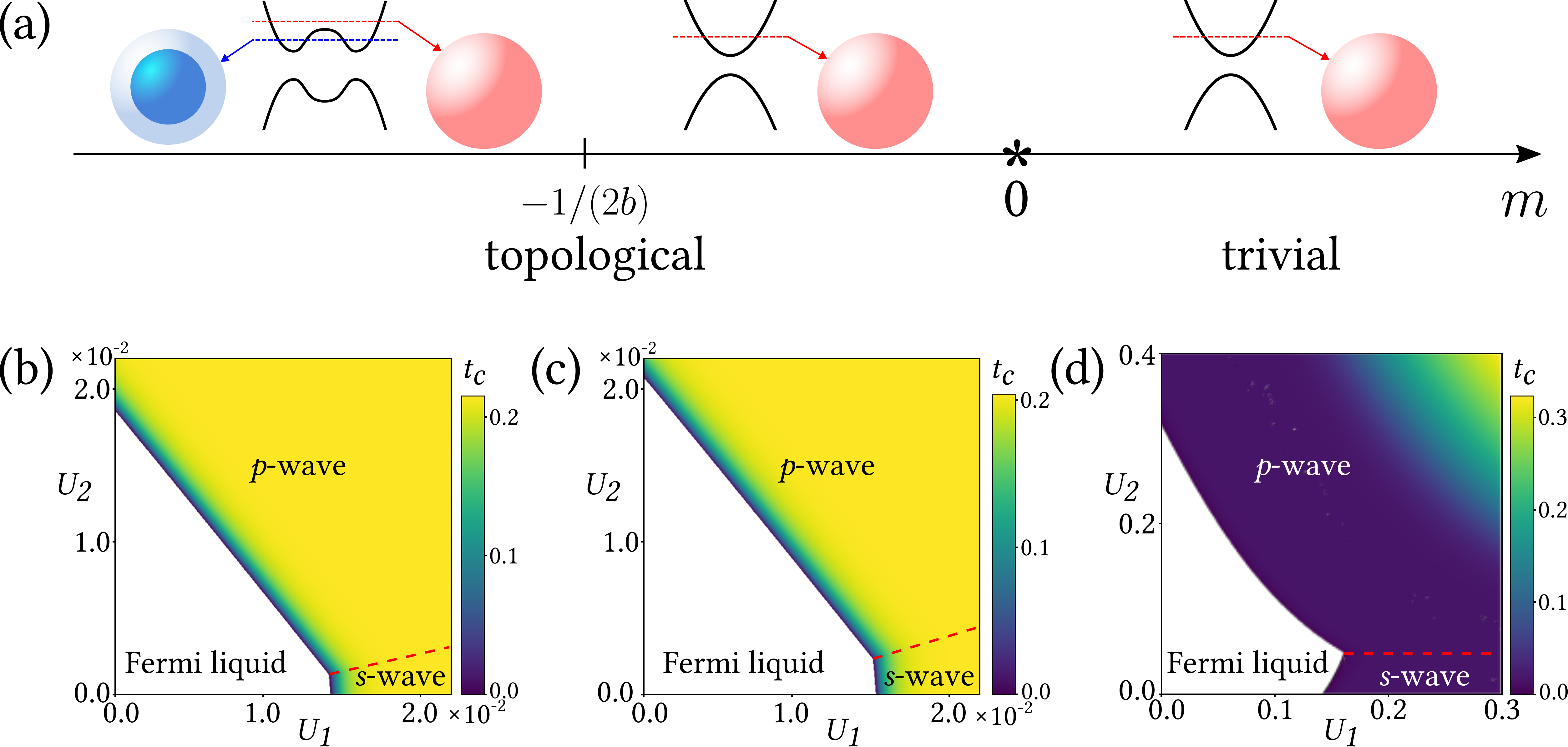}
    \caption{(a) Fermi surface (FS) geometries for various chemical doping (dashed colored lines) in a Dirac insulator in the topological ($m<0$) and trivial ($m>0$) regimes with $b>0$. A quantum phase transition between them takes place at $m=0$, around which the FS is simply connected (red spheres). Deep inside the topological regime ($m<-1/(2b)$), the system supports annular FS (concentric blue spheres). Black curves show energy eigenvalues (vertical direction) along a momentum (horizontal direction). Phase diagram in the $(U_1, U_2)$ plane for (b) $\mu=0.685$, $m=0.7$, $b=1.0$ (yielding an annular FS in doped TI), (c) $\mu=0.730$, $m=-0.7$, $b=1.0$ (yielding a simply connected FS in doped TI), (d) $\mu=0.180$, $m=0.1$, $b=0.1$ (yielding a simply connected FS in doped trivial insulator) and $t=10^{-6}$ (mimics zero temperature). Color map in (b)-(d) indicates the transition temperature ($t_c$). White regions mark a Fermi liquid without any ordering. The phase diagrams display odd-parity topological $p$-wave and trivial $s$-wave pairings, with the red dashed lines indicating the phase boundary between them.
		}~\label{fig:U1U2}
\end{figure*}

Doped Dirac insulators accommodate simply connected and annular FSs, see Fig.~\ref{fig:U1U2}(a). In this Letter, we theoretically study the effects of electron-electron ($e$-$e$) interactions on the phase diagram of such systems using an unbiased leading-order renormalization group (RG) analysis, controlled by an appropriate $\epsilon$ expansion. In strongly correlated TIs, such as the ones in Kondo systems~\cite{TKI1, TKI2, TKI3}, $e$-$e$ interactions can arise from Hubbardlike repulsions. In weakly correlated materials, effective $e$-$e$ interactions can be mediated by optical phonons below the scale of optical frequency~\cite{EPH1, EPH2, EPH3}. Without delving into their exact microscopic origin, here we consider all symmetry allowed $e$-$e$ interactions. Then within the framework of a Hubbardlike model, composed of the on-site repulsions among fermions with opposite spin ($U_1$) and parity ($U_2$) eigenvalues, we show that 3D correlated doped TIs feature an intriguing confluence of odd-parity pseudoscalar topological pairing and trivial scalar $s$-wave pairing, respectively favored by $U_2$ and $U_1$, irrespective of the underlying FS geometry. See Figs.~\ref{fig:U1U2}(b)-(d).

\emph{Model}.~The 3D massive Dirac Hamiltonian is given by
\begin{align}
\hat{h}_{\vec{k}}= \epsilon_0 + \sum_{i=1}^3 a_i k^2_i - \mu + \sum_{i=1}^3 v_i k_i \Gamma_{1i}
+ \left( m + \sum_{i=1}^3 b_i k^2_i \right) \Gamma_{30},\label{eq:Hamiltonian}
\end{align}
where $(1,2,3) \equiv (x,y,z)$. All the Hermitian matrices are represented as Kronecker products $\Gamma_{\mu \nu}=\tau_\mu \otimes \sigma_\nu$. Two sets of Pauli matrices $\tau_\mu$ and $\sigma_\mu$ act on the parity and spin indices, respectively, where $\mu=0,1,2,3$, and $\tau_0$ and $\sigma_0$ are the identity matrices. The associated spinor basis is $\Psi_{\vec{k}}^\top=[c_{+\uparrow}, c_{+\downarrow},c_{-\uparrow}, c_{-\downarrow}](\vec{k})$, where $c_{\tau \sigma}(\vec{k})$ is a fermion annihilation operator with parity $\tau=\pm$, spin $\sigma=\uparrow,\downarrow$, and momenta $\vec{k}$. Here, $v_i$ ($k_i$) is the Fermi velocity (momenta) along the $i$th coordinate and $\mu$ is the chemical potential. We consider a cubic system, for which $w_x=w_y=w_z \equiv w$ (say) for $w=v,a,b$. The Dirac Hamiltonian then enjoys the full $O(3)$ rotational symmetry, generated by $\{ \Gamma_{01}, \Gamma_{02}, \Gamma_{03} \}$. In Eq.~\eqref{eq:Hamiltonian}, $m$ ($b k^2$) describes an isotropic constant (momentum-dependent Wilson-Dirac) mass, $\epsilon_0$ ($a k^2$) captures momentum-independent (momentum-dependent) particle-hole asymmetry with $k^2 \equiv |\vec{k}|^2$. For simplicity, we set $\epsilon_0=a=0$. Then the chemical potential is measured from the band center or zero energy. The massive Dirac Hamiltonian also possesses discrete parity ($\mathcal{P}$), time reversal ($\mathcal{T}$), and charge conjugation ($\mathcal{C}$) symmetries. In the announced basis, $\mathcal{P} \Psi_{\vec{k}} \mathcal{P}=\Gamma_{30} \Psi_{-\vec{k}}$, $\mathcal{T} \Psi^\ast_{\vec{k}} \mathcal{T}=-\Gamma_{02} \Psi_{-\vec{k}}$, and $\mathcal{C} \Psi_{\vec{k}} \mathcal{C}=-i\Gamma_{12} \Psi^\ast_{\vec{k}}$~\cite{TITh0, DiracTh1, DiracTh2, DiracTh3}.

The single-particle spectra of $\hat{h}_{\vec{k}}$, hosting two distinct phases for $\mu=0$, are $\pm \sqrt{v^2 k^2 + (m+b k^2)^2}-\mu$. For $m b>0$, the system is in the trivial phase with minimal spectral gap $2m$, and setting $\mu>m$ yields a simply connected FS. On the other hand, for $m b<0$, the system is in the topological phase. Inside the topological regime, for $2 m b<-1$ the energy spectra show a ``sagging'' of the bands, thereby supporting an annular (simply connected) FS for $\sqrt{-1-4 b m}/(2b)<\mu< m$ ($\mu>m$). By contrast, when $-1<2 m b<0$, the FS in the doped TI is simply connected. From now on we set $b>0$ and tune the sign of $m$ to describe the trivial ($m>0$) and topological ($m<0$) phases. The band gap closes at a quantum critical point $m=0$, hosting massless Dirac excitations around $\vec{k}=0$.

\emph{Interactions}.~In this Letter, we investigate the superconducting instabilities in doped Dirac insulators triggered by nonretarded local ``Hubbardlike'' interactions of the form $(\Psi^\dag \Gamma_{\mu \nu} \Psi)(\Psi^\dag \Gamma_{\rho \lambda} \Psi)$. Upon taking into account the cubic or rotational, $\mathcal{P}$, $\mathcal{T}$ and $\mathcal{C}$ symmetries, the interacting Lagrangian containing all (eight) such symmetry-allowed four-fermion terms reads~\cite{DiracTh2, DiracTh3}
\begin{align}~\label{eq:allinteractions}
L_{\mathrm{int}}= \sum^3_{\beta=0} g^{\rm s}_{_\beta} (\Psi^\dag \Gamma_{\beta 0} \Psi)^2
+ \sum^3_{\beta=0} g^{\rm t}_{_\beta} \left[ \sum^3_{j=1}(\Psi^\dag \Gamma_{\beta j} \Psi)^2 \right]. 
\end{align}
Here, the superscript ${\rm s}$ (${\rm t}$) represents a singlet (triplet) in the spin sector. But, the number of linearly independent quartic couplings is reduced to four via the Fierz identity~\cite{DiracTh4}, which we choose to be the ones in the spin-singlet channels. These derivations are shown in the Supplemental Material (SM)~\cite{supplementary}. Then the imaginary time ($\tau$) action can be decomposed as $S=S_0+S_{\mathrm{int}}$, where 
\allowdisplaybreaks[4]
\begin{eqnarray}
S_0 &=&\int \D \tau \int \D^d\vec{x} \; \Psi^\dag_{\tau,\vec{x}} \left[ \partial_\tau + \hat{h}_{\vec{k}\to -i\boldsymbol{\nabla}} \right]\Psi_{\tau,\vec{x}},\\
S_{\mathrm{int}} &=& \int \D \tau \int \D^d \vec{x} \; \left(\; \sum^3_{\beta=0} g^{\rm s}_{_\beta} \; (\Psi^\dag_{\tau,\vec{x}} \Gamma_{\beta 0} \Psi_{\tau,\vec{x}})^2 \right),~\label{eq:S_int}
\end{eqnarray}
and $\Psi^\dag_{\tau,\vec{x}}$ and $\Psi_{\tau,\vec{x}}$ are independent Grassmann variables. Any microscopic model of local interactions can be captured by the vector coupling $\vec{g}^{\rm s}$. Here we consider a multi-orbital Hubbard model with only on-site density-density repulsions ($U_1, U_2>0$) with the Hamiltonian 
\begin{equation}
H^{{\rm micro}}_{{\rm int}}= U_1 n_\uparrow n_\downarrow +  U_2 n_+ n_-, 
\end{equation}
where $n_\sigma= n_{+ \sigma} + n_{- \sigma}$ for $\sigma=\uparrow, \downarrow$, $n_\tau= n_{\tau \uparrow} + n_{\tau \downarrow}$ for $\tau=\pm$, and $n_{\tau \sigma}=c^\dag_{\tau \sigma} c_{\tau \sigma}$ is the fermionic density with parity $\tau$ and spin projection $\sigma$. Single-orbital systems allow only $U_1$ (conventional on-site Hubbard repulsion), while $U_2$ is nontrivial due to the parity degrees of freedom. In terms of $\vec{g}^{\rm s}$, the above model is captured by~\cite{supplementary}
\vspace{-0.20cm}
\begin{equation}~\label{eq:U1U2intial}
    g_0^{\rm s}=U_1/2 + U_2/4, \:\:
		g_1^{\rm s}=g_2^s=U_1/8, \:\:
		g_3^{\rm s}=U_1/8-U_2/4.
\end{equation}

\begin{table}[t!]
\begin{tabular}{|c|c|c|c|c|c|}
\hline
CF & Matrix & Physical meaning & ${\mathcal P}$ & ${\mathcal T}$ & ${\mathcal C}$  \\
\hline \hline
$\Delta_0^{\rm p}$ &  $\eta_\alpha\Gamma_{00}$ & Scalar $s$-wave pairing & + & -/+ & + \\
$\Delta_1^{\rm p}$ &  $\eta_\alpha\Gamma_{10}$ & Pseudoscalar $p$-wave pairing & - & -/+ & +  \\
$\Delta_2^{\rm p}$ &  $\eta_\alpha\Gamma_{2j}$ & Spatial vector (nematic) pairing & - & -/+ & + \\
$\Delta_3^{\rm p}$ &  $\eta_\alpha\Gamma_{30}$ & Temporal vector pairing & + & -/+ & + \\
\hline \hline
$\Delta_0^{\rm s}$ &  $\eta_3\Gamma_{00}$ & Fermionic density & + & + & -  \\
$\Delta_1^{\rm s}$ &  $\eta_3\Gamma_{10}$ & Chiral density & - & + & +\\
$\Delta_2^{\rm s}$ &  $\eta_0\Gamma_{20}$ & Pseudoscalar mass & - & - & +  \\
$\Delta_3^{\rm s}$ &  $\eta_3\Gamma_{30}$ & Scalar mass & + & + & +  \\
$\Delta_0^{\rm t}$ &  $\eta_0\Gamma_{0j}$ &  Axial current & + & - & + \\
$\Delta_1^{\rm t}$ &  $\eta_0\Gamma_{1j}$ &  Abelian current & - & - & -  \\
$\Delta_2^{\rm t}$ &  $\eta_3\Gamma_{2j}$ &  Spatio-temporal tensor & - & + & - \\
$\Delta_3^{\rm t}$ &  $\eta_0\Gamma_{3j}$ & Spatial tensor & + & - & - \\
\hline
\end{tabular}
\caption{Local superconducting (excitonic) orders with their conjugate fields (CF, first column), matrices associated with the fermion bilinears $\Psi^\dag_{\rm Nam} \eta_\mu \Gamma_{\nu \rho} \Psi_{\rm Nam}$ in the Nambu basis (second column), and their physical meaning (third column) are shown in the upper (lower) block. Here, $j=1,2,3$ and in the superconducting channels $\alpha=1,2$, reflecting the $U(1)$ gauge redundancy in the choice of the superconducting phase ($\phi$) [Eq.~\eqref{eq:localOP2}]. Their transformations under the discrete ${\mathcal P}$, ${\mathcal T}$, and ${\mathcal C}$ symmetries are shown in the fourth, fifth and sixth columns, respectively, where $+$ ($-$) indicating even (odd).
}    
\label{tab:bilinears}
\end{table}

\emph{Orders}.~For a systematic analysis of superconducting instabilities, we now introduce a Nambu spinor $\Psi_{\rm Nam}(\vec{k})=(\Psi_{\vec{k}}, \Gamma_{02} \Psi^\ast_{-\vec{k}})$, where in the lower block we absorb the unitary part of the time-reversal operator. The Dirac Hamiltonian in this basis reads as $\hat{h}_{\rm Nam}(\vec{k})=\eta_3 \hat{h}_{\vec{k}}$, where $\eta_\mu$ is another set of Pauli matrices acting on the newly introduced Nambu degree of freedom. The effective action containing all symmetry-allowed local pairing (pair) and excitonic (exc) orders takes the form
\allowdisplaybreaks[4]
\begin{align}~\label{eq:localOP1}
S_{\mathrm{local}}=\int \D \tau \int \D^d \vec{r} \Psi_{\rm Nam}^\dag (\hat{h}_{\mathrm{pair}}+\hat{h}_{\mathrm{exc}}) \Psi_{\rm Nam},
\end{align}
where
\allowdisplaybreaks[4]
\begin{align}~\label{eq:localOP2}
\hat{h}_{\mathrm{pair}}&=(\eta_1 \cos \phi + \eta_2 \sin \phi)  \nonumber \\
&\times \Big[ \Delta_0^{\rm p} \Gamma_{00} + \Delta_1^{\rm p} \Gamma_{10}
+ \Delta_2^{\rm p}\sum_{j=1}^3 \Gamma_{2j} + \Delta_3^{\rm p} \Gamma_{30}  \Big],
\end{align}
with $\phi$ as the $U(1)$ superconducting phase, and
\allowdisplaybreaks[4]
\begin{align}~\label{eq:localOP3}
\hat{h}_{\mathrm{exc}}&=\Delta_0^{\rm s} \eta_3\Gamma_{00} + \Delta_1^{\rm s} \eta_3\Gamma_{10} +\Delta_2^{\rm s} \eta_0\Gamma_{20}  + \Delta_3^{\rm s} \eta_3\Gamma_{30}  \\
&+\sum_{j=1}^3 \Bigg[ \Delta_0^{\rm t} \eta_0\Gamma_{0j} +\Delta_1^{\rm t} \eta_0\Gamma_{1j} + \Delta_2^{\rm t} \eta_3\Gamma_{2j} + \Delta_3^{\rm t} \eta_0\Gamma_{3j} \Bigg]. \nonumber
\end{align}
The physical meanings and symmetry properties of all the local orders are summarized in Table~\ref{tab:bilinears}. Around the FS, the scalar (pseudoscalar) pairing takes the form of singlet $s$-wave (triplet $p$-wave) pairing~\cite{SCTh5}.

\emph{RG analysis}.~We showcase the superconducting instabilities in doped Dirac insulators, resulting from the repulsive spin ($U_1$) and orbital ($U_2$) Hubbard interactions, from a leading-order unbiased RG analysis. To this end we introduce an ultraviolet momentum cutoff $\Lambda \sim 1/a$, where $a$ is the lattice constant. We then successively lower the cutoff by integrating out thin momentum shells, thus progressively accessing the long wavelength behavior of the interacting system. The scaling dimension of momentum is $[k]=1$ and that of (Matsubara) frequency is $[\omega]=z$, where $z=1$ is the dynamical critical exponent of a Dirac system. The scale invariance of the imaginary time action $S$ implies $[\Psi^\dagger]=[\Psi]=d/2$ and $[g_\mu^\alpha]=z-d$, facilitating a controlled $\epsilon$ expansion about the lower critical one spatial dimension, where the local quartic interactions are marginal, with $\epsilon=d-1$.

Upon accounting for one-loop quantum corrections, the RG flow equations for the dimensionless quartic couplings, defined as $ 4 \pi g_\beta^{\rm s} \Lambda^\epsilon/v \to g_\beta^{\rm s}$, take the form
\begin{align}
    \frac{\D g^{\rm s}_\beta}{\D \ell} = -\epsilon g^{\rm s}_\beta + \sum_{\nu,\rho} H^\beta_{\nu \rho}(t,\mu, m, b) g^{\rm s}_\nu g^{\rm s}_\rho,
\end{align}
where $\ell$ is the logarithm of the RG scale. The functions $H^\beta_{\nu \rho}$ depend on dimensionless temperature $t=T/(\Lambda v)$, chemical potential $\mu=\mu/(\Lambda v)$, and masses $m=m/(\Lambda v)$ and $b=b \Lambda/v$. Their RG flow equations are 
\begin{align}~\label{eq:flowparameter}
    \frac{\D \mu }{ \D \ell}& = z \mu, & \frac{\D t }{ \D \ell}& = z t, &
    \frac{\D m }{ \D \ell}& = z m, & \frac{\D b }{ \D \ell}& = (z-2) b.
\end{align}
Therefore $\mu$, $t$, and $m$ are relevant parameters, while $b$ is an irrelevant quantity under coarse grain. The scale invariance of $S_{\rm local}$ implies $[\Delta_\beta^\alpha]=z$, and the corresponding RG flow equations after the one-loop quantum corrections assume the following generic form
\begin{align}
    \frac{\D \: {\rm ln}\Delta_\beta^\alpha}{\D \ell} -z= \sum_{\nu} F_{\nu}^{\beta}(t,\mu, m, b) g^{\rm s}_\nu.
\end{align}
The functions $H_{\nu \rho}^\beta$ and $F_\nu^{\beta}$ are shown in the SM~\cite{supplementary}.

In the presence of an underlying FS, RG flows of $\vec{g}^{\rm s}_\beta$, whose bare values at the scale of the ultraviolet cutoff $\Lambda$ or $\ell=0$ is set by Eq.~\eqref{eq:U1U2intial}, have to be stopped at an infrared scale $\ell^\star_\mu=-\ln(\mu(0))/z$, where $\mu(0)<1$ is the bare value of $\mu$~\cite{vafek, juricicroytwisted, szabomoessnerroy, Szabo_Roy_2021}. In the disordered phase (without any symmetry breaking) all $g_\beta^{\rm s}$s decrease under coarse grain, and none of them diverge. Its phase boundary with the ordered phase is characterized by the divergence of at least one of the coupling constants at an RG scale $\ell_{\rm div} =\ell^\star_\mu$. Within the ordered phase the respective coupling constants diverge at an RG scale $\ell_{\rm div}<\ell^\star_\mu$. The exact nature of the leading instability or the pattern of symmetry breaking is then unambiguously identified by the renormalized susceptibility $\Delta_\beta^\alpha$ that receives the largest positive correction (anomalous dimension) when one of the coupling constants diverges at the RG scale $\ell_{\rm div}$.

Throughout, we set $t =10^{-6}$ to construct the zero-temperature phase diagrams. Nonetheless, the notion of $\ell_{\rm div}$ provides a poor man's estimation of the transition temperature ($t_c$) in the following way. The solution of the RG flow equation for dimensionless temperature from Eq.~\eqref{eq:flowparameter} reads as $t(\ell)=t(0)\exp(z \ell)$. In an interacting doped Dirac insulator, there are two infrared RG scales $\ell^\star_\mu$ and $\ell_{\rm div}$, respectively at which the renormalized chemical potential and at least one of the coupling constants becomes of the order of \emph{unity}. Associated with these two RG scales, we can define two bare temperatures $t(0)=t_\mu$ and $t_g$, respectively, such that $t(\ell^\star_\mu)=t(\ell_{\rm div})=1$, yielding $t_\mu=\exp[-z \ell^\star_\mu]$ and $t_g=\exp[-z \ell^\star_{\rm div}]$. The transition temperature in the ordered phase is then defined as $t_c=t_g - t_\mu$ (up to an overall numerical prefactor). Defined this way, $t_c=0$ at the phase boundary between the disordered and ordered (superconducting) phases, and inside the ordered phase $t_c$ increases monotonically with the increasing coupling strength, as then $\ell_{\rm div}$ decreases monotonically.

\renewcommand*{\arraystretch}{1.3}
\begin{table}[]
    \centering
    \begin{tabular}{|c|c |c|c| c|c | c|c|}
    \hline

      \multirow{2}{*}{CC} &  \multicolumn{3}{c|}{Excitonic order} & \multicolumn{3}{c|}{Pairing order} & ES \\
      \cline{2-7}
            & CF & Symmetry & SR & CF & Symmetry & SR & SR3  \\
      \hline
      $g_0^{\rm s}$ &  -- &  -- & -- & $\Delta^{\rm p}_0$ & $O(2)$ & SR2 &-- \\
      $g_1^{\rm s}$ & $\Delta^{\rm s}_1$ & $O(1)$/$Z_2$ & SR1 & $\Delta^{\rm p}_0$ & $O(2)$ & SR2 & $O(3)$ \\
      $g_2^{\rm s}$ & $\Delta^{\rm s}_2$ & $O(1)$/$Z_2$ & SR1 & $\Delta^{\rm p}_1$ & $O(2)$ & SR2 & $O(3)$ \\
      $g_3^{\rm s}$ & $\Delta^{\rm s}_3$ & $O(1)$/$Z_2$ & SR1 & $\Delta^{\rm p}_0$ & $O(2)$ & SR2 & $O(3)$ \\
      $g_0^{\rm t}$ & $\Delta^{\rm t}_0$ & $O(3)$ & SR1 & $\Delta^{\rm p}_2$ & $O(3)$ & SR2 & $O(3)$ \\
      $g_1^{\rm t}$ & $\Delta^{\rm t}_1$ & $O(3)$ & SR1 & $\Delta^{\rm p}_3$ & $O(2)$ & SR2 & $O(5)$ \\
      $g_2^{\rm t}$ &   -- & -- & -- & $\Delta^{\rm p}_0$ & $O(2)$ & SR2 & -- \\
      $g_3^{\rm t}$ & $\Delta^{\rm s}_2$ & $O(1)$/$Z_2$ & SR2 & $\Delta^{\rm p}_1$ & $O(2)$ & SR2 & $O(3)$ \\
      \hline
    \end{tabular}
    \caption{Excitonic and pairing instabilities, indicated by their respective conjugate fields (CFs), for individual repulsive interactions with coupling constant (CC) $g_\beta^\alpha>0$, with all other CCs set to zero. We highlight the role of the selection rule (SR) for each such order (see text), along with the symmetry of the individual order parameters. In the right most column we indicate the enlarged symmetry (ES) of the composite order parameter, constructed by combining those for the excitonic and pairing orders, following the SR3. Dashed lines indicate the absence of any excitonic ordering and hence any ES of the composite order parameter.  
		}~\label{tab:individual_channels}
\end{table}

\emph{Phase diagram}.~We follow the outlined procedure to construct the phase diagram for the interacting doped 3D Dirac insulator in the $(U_1,U_2)$ plane for different underlying FS topologies [Fig.~\ref{fig:U1U2}(a)]. The results are shown in Figs.~\ref{fig:U1U2}(b)-(d). While repulsive $U_1$ favors nucleation of conventional scalar $s$-wave pairing, repulsive $U_2$ is conducive to the condensation for the odd-parity pseudoscalar $p$-wave pairing. As the relative strength of these two interactions is varied, there is a transition between them, around which, in principle, the system can also foster an axionic $p+is$ paired state~\cite{SCTh4}, which, however, cannot be captured from the RG analysis. These outcomes are insensitive to the underlying FS topology.

To gain insights into the nature of the superconducting ground states in the $(U_1,U_2)$ plane, it is worth focusing on the $U_1$ and $U_2$ axes separately. As the strength of repulsive $U_1$ is increased beyond its critical value, $g^{\rm s}_{0,1,3} \to + \infty$ ($g^{\rm s}_2 \to -\infty$), becoming effectively repulsive (attractive). From the unbiased RG procedure it can be shown that bare repulsive $g^{\rm s}_{0,1,3}$ and bare attractive $g^{\rm s}_2$ favor $s$-wave pairing, as also found by tuning solely $U_1$. On the other hand, when the strength of $U_2$ is tuned beyond the critical one, $g^{\rm s}_{0,1,2} \to + \infty$ while $g^{\rm s}_{3} \to - \infty$. RG flow with bare repulsive $g^{\rm s}_{0,1}$ favors $s$-wave pairing, whereas bare repulsive (attractive) $g^{\rm s}_{2}$ ($g^{\rm s}_{3}$) favors pseudoscalar $p$-wave pairing. When the strength of only $U_2$ is tuned, the degree of divergence for $g^{\rm s}_{2,3}$ is larger than that for the $g^{\rm s}_{0,1}$, causing condensation of odd-parity pseudoscalar pairing. These observation can be further substantiated from recently proposed ``selection rules" (SRs) between the dominant interaction and the resulting ordered state(s)~\cite{juricicroytwisted, szabomoessnerroy, Szabo_Roy_2021}, which we discuss next.

\emph{Selection rules}.~We consider a single quartic term and a fermion bilinear, $g_Q\sum_i (\Psi^\dag_{\rm Nam} Q_i \Psi_{\rm Nam})^2$ and $\sum_i \Psi^\dag_{\rm Nam} R_i \Psi_{\rm Nam}$, respectively, chosen from the sets of all the symmetry-allowed four-fermion interactions [Eq.~\eqref{eq:allinteractions}] and order parameters [Eqs.~\eqref{eq:localOP1}-\eqref{eq:localOP3}]. Here $Q_i$ and $R_i$ are Hermitian matrices, expressed in the announced Nambu basis. Let $A_M$ be the number of anticommuting matrix pairs between the four-fermion interaction and order parameter terms. For concreteness, we consider all the interactions to be repulsive ($g_Q>0$). Then among all the available ordered phases, the interaction channel $g_Q$ maximally boosts the nucleation of the ones for which $Q_i \equiv R_i$ (SR1) or $A_M$ is maximal (SR2). It is also conceivable for a given repulsive interaction to support excitonic (for zero and low $\mu$) and superconducting (for moderate $\mu$) states, described by the fermion bilinears $\sum^{K}_{i=1} \Psi^\dag_{\rm Nam} R^{\rm exc}_i \Psi_{\rm Nam}$ and $\sum^{L}_{i=1} \Psi^\dag_{\rm Nam} R^{\rm pair}_i \Psi_{\rm Nam}$, respectively, transforming as vectors under $O(K)$ and $O(L)$ rotations, constituted by $K$ and $L$ number of mutually anticommuting matrices, respectively. Then these two ordered phases can appear next to each other when, for example, $\mu$ is tuned in the system, if the corresponding vector order parameters form composite vectors that transform under the $O(N)$ rotations, where $K,L < N \leq K+L$ (SR3). These SRs have been successfully employed for twisted bilayer graphene~\cite{juricicroytwisted}, 3D (un)doped Luttinger materials~\cite{szabomoessnerroy}, and (un)doped monolayer and bilayer graphene~\cite{Szabo_Roy_2021}.

We exemplify such a seemingly mathematical formulation by considering a specific (repulsive) interaction $g^{\rm s}_2$ in doped Dirac insulator. For low chemical doping, it supports nucleation of the excitonic pseudoscalar mass, transforming as an O(1) or a $Z_2$ vector, following the SR1. In the moderate doping regime, the same interaction favors condensation of the pseudoscalar pairing, transforming as an O(2) vector, following the SR2. In the $(\mu,t)$ plane of the phase diagram with a fixed $g^{\rm s}_2>0$, shown in the SM~\cite{supplementary}, these ordered phases live next to each other, as they form a composite O(3) vector~\cite{roygoswami_Z2}, following the SR3. These outcomes qualitatively justify the appearance of the pseudoscalar pairing along the $U_2$ axis [Figs.~\ref{fig:U1U2}(b)-(d)]. Similar conclusions hold when we tune any one of the eight local four-fermion interactions [Eq.~\eqref{eq:allinteractions}]. The results are summarized in Table~\ref{tab:individual_channels}, and the corresponding cuts of the phase diagrams in the $(\mu,t)$ plane for fixed values of $g^{{\rm s}/{\rm t}}_\beta>0$ are shown in the SM~\cite{supplementary}.

Notice that emergence of the superconductivity from purely repulsive electron-electron interactions in doped TIs, featuring a FS, follows the spirit of the Kohn-Luttinger mechanism~\cite{KL1, KL2, KL3}. The presence of a FS suppresses the interband scattering, responsible for particle-hole or excitonic orderings, while the intraband scattering remains unaffected yielding superconducting ground states, shown in the phase diagram of the Hubbard model in Fig.~\ref{fig:U1U2}(b), for example. Nonetheless, as summarized in Table~\ref{tab:individual_channels}, repulsive electronic interactions also accommodate various excitonic orders, mentioned in Table~\ref{tab:bilinears}, especially in the absence of an underlying FS~\cite{supplementary}. Here we support these generic outcome from a controlled and unbiased RG analysis. As there is no Fermi-surface nesting in doped TIs, no charge- or spin-density-wave ordering can develop via nesting, and all the particle-hole or excitonic orderings result from inter-band scattering, which gets suppressed by chemical doping.

\emph{Summary and discussion}.~Here, we formulate an unbiased and controlled RG framework to capture dominant superconducting instabilities of 3D doped interacting Dirac insulators (topological and trivial), featuring an intriguing confluence of topological $p$-wave and trivial $s$-wave pairings, triggered by on-site Hubbard repulsions. Our approach is sufficiently general to address similar phenomenon in doped TIs belonging to any Altland-Zirnbauer symmetry class in any dimension $d>1$, with special emphasis on two-dimensional doped TIs~\cite{thomale, ytHsu, schnyder}, as well as in doped crystalline TIs, harboring FSs near multiple high symmetry points in the Brillouin zone, connected by discrete crystal symmetries, and in doped higher-order TIs and topological Dirac and Weyl semimetals. The present analysis can be generalized to encompass longer, but finite-range Coulomb repulsions as well as layered topological materials possessing tetragonal symmetry, for which $w_x=w_y\neq w_z$, where $w=v,b,a$. These generalizations may allow the appearance of other competing paired states (such as the nematic one), tabulated in Table~\ref{tab:bilinears}, in their global phase diagram. The current work lays the theoretical foundation for such fascinating future research directions.

\emph{Acknowledgments}.~A.S.\ is grateful for financial support from the Swiss National Science Foundation (SNSF) through Division II (No.~184739). B.R.\ was supported by NSF CAREER Grant No.\ DMR-2238679, and is thankful to Vladimir Juri\v{c}i\'c and Sanjib Kumar Das for a critical reading of the manuscript.


\begin{thebibliography}{}


\bibitem{TITh0} M. E. Peskin and D. V. Schroeder, \emph{An introduction to quantum field theory} (CRC Press, Boca Raton, FL, 2019).

\bibitem{TITh1} M. Z. Hasan and C. L. Kane, Colloquium: Topological insulators, Rev.\ Mod.\ Phys.\ {\bf 82}, 3045 (2010).

\bibitem{TITh2} X.-L. Qi and S.-C. Zhang, Topological insulators and superconductors, Rev.\ Mod.\ Phys.\ {\bf 83}, 1057 (2011).

\bibitem{TITh3} C.-K. Chiu, J. C. Y. Teo, A. P. Schnyder, and S. Ryu, Classification of topological quantum matter with symmetries, Rev.\ Mod.\ Phys.\ {\bf 88}, 035005 (2016).

\bibitem{TITh4} A. Bansil, H. Lin, and T. Das, Colloquium: Topological band theory, Rev.\ Mod.\ Phys.\ {\bf 88}, 021004 (2016).

\bibitem{TITh5} A. Altland and M. R. Zirnbauer, Nonstandard symmetry classes in mesoscopic normal-superconducting hybrid structures, Phys.\ Rev.\ B {\bf 55}, 1142 (1997).

\bibitem{TITh6} C. L. Kane and E. J. Mele, ${Z}_{2}$ Topological Order and the Quantum Spin Hall Effect, Phys.\ Rev.\ Lett.\ {\bf 95}, 146802 (2005).

\bibitem{TITh7} B. A. Bernevig, T. L. Hughes, and S.-C. Zhang, Quantum Spin Hall Effect and Topological Phase Transition in HgTe Quantum Wells, Science {\bf 314}, 1757 (2006).

\bibitem{TITh8} L. Fu and C. L. Kane, Topological insulators with inversion symmetry, Phys.\ Rev.\ B {\bf 76}, 045302 (2007).

\bibitem{TITh9} A. Kitaev, Periodic table for topological insulators and superconductors, AIP Conf.\ Proc.\ {\bf 1134}, 22 (2009).

\bibitem{TITh10} C.-X. Liu, X.-L. Qi, H. Zhang, X. Dai, Z. Fang, and S.-C. Zhang, Model Hamiltonian for topological insulators, Phys.\ Rev.\ B {\bf 82}, 045122 (2010).

\bibitem{TITh11} S. Ryu, A. P. Schnyder, A. Furusaki, and A. W. W. Ludwig, Topological insulators and superconductors: tenfold way and dimensional hierarchy, New J.\ Phys.\ {\bf 12}, 065010 (2010).

\bibitem{TITh12} G. E. Volovik, \emph{The Universe in a Helium Droplet} (Oxford University Press, Oxford, UK, 2009).

\bibitem{TITh13} S.-Q. Shen, \emph{Topological Insulators: Dirac Equation in Condensed Matter}, 2nd ed.\ (Springer, Singapore, 2017).

\bibitem{TITh14} B. A. Bernevig and T. L. Hughes, \emph{Topological Insulators and Topological Superconductors} (Princeton University Press, NJ, 2013).

\bibitem{TITh15} R. Dornhaus, G. Nimtz, and B. Schlicht, \emph{Narrow-Gap Semicounductors} (Springer-Verlag, Berlin, 1983).


\bibitem{TSC:Exp1} A. L. Wray, S-Y. Xu, Y. Xia, Y. S. Hor, D. Qian, A. V. Fedorov, H. Lin, A. Bansil, R. J. Cava, and M. Z. Hasan, Observation of topological order in a superconducting doped topological insulator, Nat.\ Phys.\ {\bf 6}, 855 (2010).

\bibitem{TSC:Exp2} M. Kriener, K. Segawa, Z. Ren, S. Sasaki, and Y. Ando, Bulk Superconducting Phase with a Full Energy Gap in the Doped Topological Insulator ${\mathrm{Cu}}_{x}{\mathrm{Bi}}_{2}{\mathrm{Se}}_{3}$, Phys.\ Rev.\ Lett.\ {\bf 106}, 127004 (2011).

\bibitem{TSC:Exp3} S. Sasaki, Z. Ren, A. A. Taskin, K. Segawa, L. Fu, and Y. Ando, Odd-Parity Pairing and Topological Superconductivity in a Strongly Spin-Orbit Coupled Semiconductor, Phys.\ Rev.\ Lett.\ {\bf 109}, 217004 (2012).

\bibitem{TSC:Exp4} T. Kirzhner, E. Lahoud, K. B. Chaska, Z. Salman, and A. Kanigel, Point-contact spectroscopy of Cu${}_{0.2}$Bi${}_{2}$Se${}_{3}$ single crystals, Phys.\ Rev.\ B {\bf 86}, 064517 (2012).

\bibitem{TSC:Exp5} N. Levy, T. Zhang, J. Ha, F. Sharifi, A. A. Talin, Y. Kuk, and J. A. Stroscio, Experimental Evidence for $s$-Wave Pairing Symmetry in Superconducting ${\mathrm{Cu}}_{x}{\mathrm{Bi}}_{2}{\mathrm{Se}}_{3}$ Single Crystals Using a Scanning Tunneling Microscope, Phys.\ Rev.\ Lett.\ {\bf 110}, 117001 (2013).

\bibitem{TSC:Exp6} M. Novak, S. Sasaki, M. Kriener, K. Segawa, and Y. Ando, Unusual nature of fully gapped superconductivity in In-doped SnTe, Phys.\ Rev.\ B {\bf 88}, 140502(R) (2013). 

\bibitem{TSC:Exp7} K. Matano, M. Kriener, K. Segawa, Y. Ando, and G-Q. Zheng, Spin-rotation symmetry breaking in the superconducting state of Cu$_x$Bi$_2$Se$_3$, Nat.\ Phys.\ {\bf 12}, 852 (2016).

\bibitem{TSC:Exp8} S. Yonezawa, K. Tajiri, S. Nakata, Y. Nagai, Z. Wang, K. Segawa, Y. Ando, and Y. Maeno, Thermodynamic evidence for nematic superconductivity in Cu$_x$Bi$_2$Se$_3$, Nat.\ Phys.\ {\bf 13}, 123 (2017).

\bibitem{TSC:Exp9} T. Asaba, B. J. Lawson, C. Tinsman, L. Chen, P. Corbae, G. Li, Y. Qiu, Y.S. Hor, L. Fu, and Lu Li, Rotational Symmetry Breaking in a Trigonal Superconductor Nb-doped ${\mathrm{Bi}}_{2}{\mathrm{Se}}_{3}$, Phys.\ Rev.\ X {\bf 7}, 011009 (2017).

\bibitem{TSC:Exp10} Y. Pan, A.M. Nikitin, G.K. Araizi, Y.K. Huang, Y. Matsushita, T. Naka, and A. de Visser, Rotational symmetry breaking in the topological superconductor Sr$_x$Bi$_2$Se$_3$probed by upper-critical field experiments, Sci.\ Rep.\ {\bf 6}, 28632 (2016).


\bibitem{SCTh1} L. Fu and E. Berg, Odd-Parity Topological Superconductors: Theory and Application to Cu$_x$Bi$_2$Se$_3$, Phys.\ Rev.\ Lett.\ {\bf 105}, 097001 (2010).

\bibitem{SCTh2} L. Fu, Odd-parity topological superconductor with nematic order: Application to Cu$_x$Bi$_2$Se$_3$, Phys.\ Rev.\ B {\bf 90}, 100509(R) (2014).

\bibitem{SCTh3} J. W. F. Venderbos, V. Kozii, and L. Fu, Odd-parity superconductors with two-component order parameters: Nematic and chiral, full gap, and Majorana node, Phys.\ Rev.\ B {\bf 94}, 180504(R) (2016).

\bibitem{SCTh4} P. Goswami and B. Roy, Axionic superconductivity in three-dimensional doped narrow-gap semiconductors, Phys.\ Rev.\ B {\bf 90}, 041301(R) (2014).

\bibitem{SCTh5} B. Roy, Y. Alavirad, and J. D. Sau, Global Phase Diagram of a Three-Dimensional Dirty Topological Superconductor, Phys.\ Rev.\ Lett.\ {\bf 118}, 227002 (2017). 



\bibitem{TKI1} M. Dzero, K. Sun, V. Galitski, and P. Coleman, Topological Kondo Insulators, Phys.\ Rev.\ Lett.\ {\bf 104}, 106408 (2010).

\bibitem{TKI2} B. Roy, J. Hofmann, V. Stanev, J. D. Sau, and V. Galitski, Excitonic and nematic instabilities on the surface of topological Kondo insulators, Phys.\ Rev.\ B {\bf 92}, 245431 (2015).

\bibitem{TKI3} M. Dzero, J. Xia, V. Galitski, and P. Coleman,  Topological Kondo Insulators , Annu.\ Rev.\ Condens.\ Matter Phys.\ {\bf 7}, 249 (2016).


\bibitem{EPH1} H. Takayama, Y. R. Lin-Liu, and K. Maki, Continuum model for solitons in polyacetylene, Phys.\ Rev.\ B {\bf 21}, 2388 (1980).

\bibitem{EPH2} G. Gr\"uner, The dynamics of charge-density waves, Rev.\ Mod.\ Phys.\ {\bf 60}, 1129 (1988).

\bibitem{EPH3} A. J. Heeger, S. Kivelson, J. R. Schrieffer, and W.-P. Su, Solitons in conducting polymers, Rev.\ Mod.\ Phys.\ {\bf 60}, 781 (1988).


\bibitem{DiracTh1} P. Goswami and S. Chakravarty, Quantum Criticality between Topological and Band Insulators in $3+1$ Dimensions, Phys.\ Rev.\ Lett.\ {\bf 107}, 196803 (2011).

\bibitem{DiracTh2} B. Roy and S. Das Sarma, Quantum phases of interacting electrons in three-dimensional dirty Dirac semimetals, Phys.\ Rev.\ B {\bf 94}, 115137 (2016).

\bibitem{DiracTh3} A. L. Szab\'o and B. Roy, Emergent chiral symmetry in a three-dimensional interacting Dirac liquid, J.\ High Energ.\ Phys.\ {\bf 2021}, 4 (2021).

\bibitem{DiracTh4} I. F. Herbut, V. Juri\v{c}i\'c, and B. Roy, Theory of interacting electrons on the honeycomb lattice, Phys. Rev. B {\bf 79}, 085116 (2009).

\bibitem{supplementary} See Supplemental Material at XXX-XXXXXX for the details of Fierz reduction and RG analysis, and phase diagrams for individual interaction channels. 



\bibitem{vafek} O. Vafek, J. M. Murray, and V. Cvetkovic, Superconductivity on the Brink of Spin-Charge Order in a Doped Honeycomb Bilayer, Phys.\ Rev.\ Lett.\ {\bf 112}, 147002 (2014).

\bibitem{juricicroytwisted} B. Roy and V. Juri\v{c}i\'c, Unconventional superconductivity in nearly flat bands in twisted bilayer graphene, Phys.\ Rev.\ B {\bf 99}, 121407(R) (2019).

\bibitem{szabomoessnerroy} A. L. Szab\'o, R. Moessner, and B. Roy, Interacting spin-$\frac{3}{2}$ fermions in a Luttinger semimetal: Competing phases and their selection in the global phase diagram, Phys.\ Rev.\ B {\bf 103}, 165139 (2021).

\bibitem{Szabo_Roy_2021} A. L. Szab\'o and B. Roy, Extended Hubbard model in undoped and doped monolayer and bilayer graphene: Selection rules and organizing principle among competing orders, Phys.\ Rev.\ B {\bf 103}, 205135 (2021).


\bibitem{roygoswami_Z2} B. Roy and P. Goswami, ${Z}_{2}$ index for gapless fermionic modes in the vortex core of three-dimensional paired Dirac fermions, Phys.\ Rev.\ B {\bf 89}, 144507 (2014).


\bibitem{KL1} W. Kohn and J. M. Luttinger, New Mechanism for Superconductivity, Phys.\ Rev.\ Lett.\ {\bf 15}, 524 (1965).

\bibitem{KL2} M. A. Baranov, A. V. Chubukov, and M. Yu. Kagan, Superconductivity and superfluidity in Fermi systems with repulsive interactions, Int.\ J.\ Mod.\ Phys.\ B {\bf 06}, 2471 (1992).

\bibitem{KL3} R. Shankar, Renormalization-group approach to interacting fermions, Rev.\ Mod.\ Phys.\ {\bf 66}, 129 (1994).


\bibitem{thomale} X. Wu, M. Fink, W. Hanke, R. Thomale, and D. Di Sante, Unconventional superconductivity in a doped quantum spin Hall insulator, Phys.\ Rev.\ B {\bf 100}, 041117(R) (2019).

\bibitem{ytHsu} Y-T. Hsu, W. S. Cole, R-X. Zhang, and J. D. Sau, Inversion-Protected Higher-Order Topological Superconductivity in Monolayer WTe$_2$, Phys.\ Rev.\ Lett.\ {\bf 125}, 097001 (2020).

\bibitem{schnyder} P. M. Bonetti, D. Chakraborty, X. Wu, and A. P. Schnyder, Interaction-driven first-order and higher-order topological superconductivity, Phys.\ Rev.\ B {\bf 109}, L180509 (2024).

\end{thebibliography}
\end{document}